 \documentclass[prl,floats,showpacs,onecolum,twocolumn]{revtex4-1}
\usepackage[draftl]{changes}
\definechangesauthor[color=blue]{IB}
\definechangesauthor[color=orange]{GB}
\definechangesauthor[color=cyan]{RS}

\usepackage[english]{babel}
\usepackage{amsmath}
\usepackage{amssymb}
\usepackage[]{graphicx}
\usepackage{times}
\usepackage{bm}
\usepackage{units}
\usepackage{color}

\begin{document}

\title{Optical fingerprint of non-covalently functionalized transition metal dichalcogenides}
\author{Maja Feierabend$^1$, Ermin Malic$^1$,  Andreas Knorr$^2$ and Gunnar Bergh\"auser$^1$}
\address{$^1$Chalmers University of Technology, Department of 
Physics, SE-412 96 Gothenburg, Sweden}
\address{$^2$Institut f\"ur Theoretische Physik, Nichtlineare Optik und Quantenelektronik, Technische Universit\"at Berlin, Hardenbergstr. 36, 10623 Berlin, Germany}

\begin{abstract}
Atomically thin transition metal dichalcogenides (TMDs) hold promising potential for applications in optoelectronics. Due to their direct band gap and the extraordinarily strong Coulomb interaction, TMDs exhibit efficient light-matter coupling and tightly bound excitons. Moreover,  large spin orbit coupling in combination with circular dichroism allows for spin and valley selective optical excitation. As atomically thin materials, they are very sensitive to changes in the surrounding environment. This motivates a functionalization approach, where external molecules are adsorbed to the materials surface to tailor its optical properties.
Here, we apply the density matrix theory to investigate the potential of non-covalently functionalized TMDs. Considering exemplary spiropyran molecules with a strong dipole moment, we predict spectral redshifts and the appearance of an additional side peak in the absorption spectrum of functionalized TMDs. 
We show that the molecular characteristics, e.g. coverage, orientation and dipole moment, crucially influence the optical properties of TMDs,  leaving a unique optical fingerprint in the absorption spectrum. 
Furthermore, we find that the molecular dipole moments open a channel for coherent intervalley coupling between the high-symmetry K and K' points which may open new possibilities for spin-valleytronics application. 

\end{abstract}

\maketitle
\section{Introduction}
The adsorption of molecules to the surface of nanostructures opens new possibilities to tune optical and electronic properties of nanomaterials \cite{pss_gunnar,gunnar_carbon,ermin_cm, ermin_prl, malic2014forster}. One especially interesting class of nanostructures are atomically thin transition metal dichalcogenides (TMDs) because they combine the ultrathin structure with an optical band gap in the range of a 1-2 eV \cite{Butler2013,Kochbuch,andor,THeinz}. Furthermore, they show a valley and spin selective polarization \cite{Cao2012,valley_pola,valley_pola2} and are therefore promising candidates for future application in valley- and spintronics. Due to the strong Coulomb and light-matter interaction the absorption spectrum shows clearly pronounced excitonic peaks \cite{RIS_0,gunnar_prb}. In addition, the atomically thin structure is very sensitive to changes in the environment, thus the functionalization with molecules could be very promising for technical devices, such as chemical sensors. \\
Here we consider TMDs non-covalently functionalized with exemplary spiropyran molecules which exhibit a large permanent dipole moment  (Fig. \ref{schema}). The structure of these molecules can be reversibly switched by an external light source between an open merocyanine and closed spiropyran ring form \cite{FisherSpiro}, leading to significant changes in the dipole moment \cite{ermin_cm}. Moreover, the spacial orientation of the molecules as well as the molecular coverage influence the interaction with the excitons in the nanostructure.
The aim of our study is to understand the optical properties of functionalized TMDs on a microscopic level. In contrast to our previous work \cite{maja_sensor} where the focus was the activation of dark $K\Lambda$ excitons by attaching molecules as a new sensing mechanism, we want to focus in this work on bright $KK$ excitons and address the question if intervalley coupling can be enabled by the molecules, cf. Fig. \ref{schema}(c).
In particular, we will estimate the influence of the molecular characteristics on the absorption spectrum of the TMDs taking  into account coupling processes between the K and K' valley, which are especially important in non-linear optics.

\section{Theoretical approach}
\begin{figure}[t]
  \begin{center}
\includegraphics[width=\linewidth]{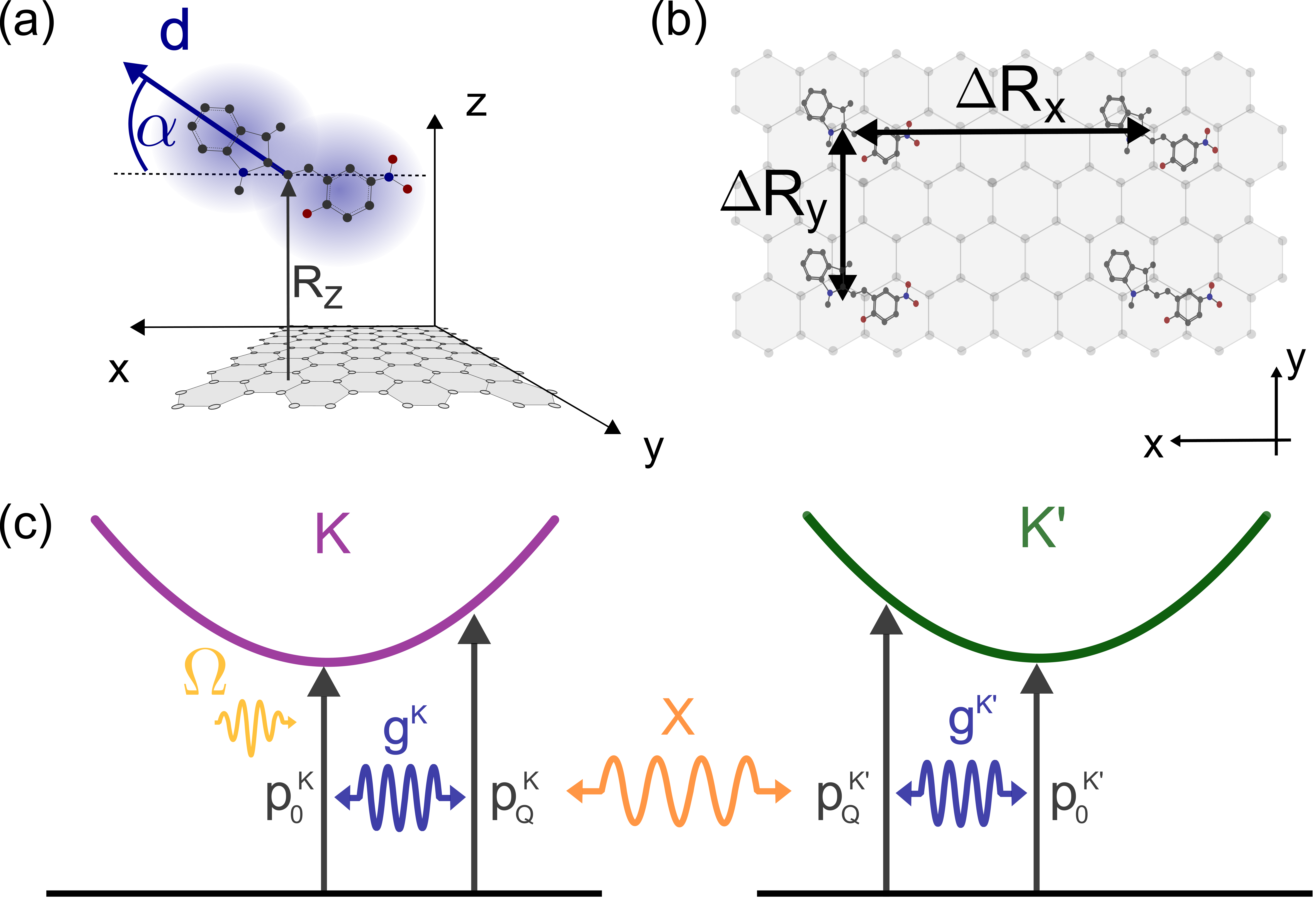} 
\end{center}
    \caption{ (a) Close up of a randomly orientated molecule attached to the TMD structure in distance $R_z \approx 0.36$ nm corresponding to the van der Waals diameter. The molecules are characterized by a dipole vector $\bf d$ inducing interaction with excitons in the TMD material. The dipole vector is represented by dipole moment $d$ and orientation along z-axis $\alpha$. The orientation within xy plane is set to zero. (b) The molecules are ordered periodically with the  lattice constant $\Delta R_x$ in x direction and $\Delta R_y$ in y direction. 
    (c) Excitonic dispersion of K and K' valley.
    The external electromagnetic field drives the excitonic coherence $p_{\mathbf Q=0}^{K}$. This gain a center of mass momentum coupling to the dipoles of the molecules (molecule-TMD coupling $g$) resulting in $p_{Q\neq0}^{K}$ which couples to  $p_{Q\neq0}^{K^\prime}$ via the Coulomb-induced intervalley coupling $X$.    
}
  \label{schema}
\end{figure}
\begin{figure}[t]
  \begin{center}
    \includegraphics[width=\linewidth]{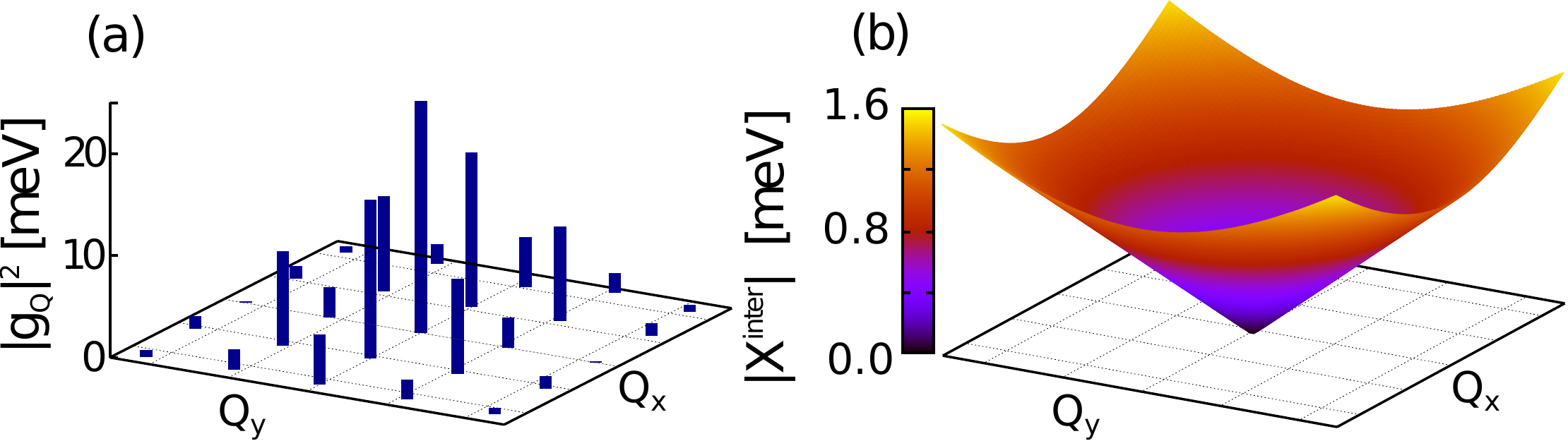}
  \end{center}
  \caption{(a) The molecule-TMD coupling (MTC) element and (b) intervalley-coupling (IVC) element for the allowed center of mass momentum $\bf Q$. Due to the dumbbell-like shape of the dipole potential (cf. \eqref{dipolfield}) the MTC element exhibits a maximum for $\bf Q=0$ and decreases with $\bf Q$. Depending on the orientation of the dipole vector within the xy plane, contributions in x and y direction are more and less pronounced. In contrast the IVC element is zero for $\bf Q=0$ due to the fact that the system needs a center of mass momentum to enable intervalley coupling processes. For increasing $\bf Q$ the IVC increases.}
 \label{kopplungselemente}
\end{figure}

In this study we focus on the optical properties of bright (optically accessible) $A_{\rm 1s}$ excitons in functionalized MoS$_2$. The absorption coefficient $\alpha(\omega)$ \cite{Kochbuch} is proportional to the optical susceptibility $\chi(\omega)$ as the linear response to an optical perturbation induced by an external vector potential $ A(\omega)$ 
\begin{equation}\label{Elliott}
\alpha(\omega) \propto \omega \Im\left[\chi(\omega)\right] \propto \frac{\Im\left[ \sum_{\bf i \bf j} M_{\bf i \bf j}^{}  p_{\bf i \bf j}^{}(\omega) + cc \right ]}{\omega A(\omega)}
\end{equation}
and can be calculated via the microscopic polarization \mbox{$p_{\bf{ij}}= \langle a_{\bf i}^{\dagger} a_{\bf j}^{}\rangle$} with compound index $\bf {i}$ including momentum $k_{i}$, band $\lambda_{i}$ and spin $\sigma_{i}$, and the optical matrix element $M_{\bf ij}^{}$. To obtain the temporal evolution of $p_{\bf{ij}}$ we solve the Heisenberg equation of motion $i\hbar \dot p_{\bf{ij}}(t)=[p_{\bf{ij}},H]$ \cite{Kochbuch, carbonbuch}. 
Here we use the many-particle Hamiltonian $H$ which includes 
the free carrier contribution $H_0=\sum_{\boldsymbol{l}} \epsilon_{\boldsymbol{l}}^{} a_{\boldsymbol{l}}^{\dagger}a _{\boldsymbol{l}}^{}$ with the electronic band structure $\epsilon_{\boldsymbol{l}}^{}$ \cite{Xiao,ochoa}, 
the carrier-light interaction 
$H_{c-l}=\frac{i\hslash e_{0}}{m_{0}} \sum_{\boldsymbol{l_{1}l_{2}}} \boldsymbol {M_{\boldsymbol{l_{1}l_{2}}}^{}} \boldsymbol{A}(t)a_{\boldsymbol{l_{1}}}^{\dagger}a _{\boldsymbol{l_{2}}}^{} $ with the electron mass $m_0$ and the elementary charge $e_{0}$,
the carrier-carrier interaction 
$H_{c-c}=\frac{1}{2}\sum_{\boldsymbol{l_{1}l_{2}l_{3}l_{4}}} V_{\boldsymbol{l_{3}l_{4}}} ^{\boldsymbol{l_{1}l_{2}}} a_{\boldsymbol{l_{1}}}^{\dagger} a_{\boldsymbol{l_{2}}}^{\dagger} a_{\boldsymbol{l_{4}}}^{}a_{\boldsymbol{l_{3}}}^{}$ 
and
the carrier-molecule interaction 
$H_{c-m}=\sum_{\boldsymbol{l_{1}l_{2}}} g_{\boldsymbol{l_{1}l_{2}}}^{} a_{\boldsymbol{l_{1}}}^{\dagger}a _{\boldsymbol{l_{2}}}^{}
$ with molecule-TMD coupling (MTC) element $g_{\boldsymbol{l_{1}l_{2}}}^{}= \langle \psi_{\boldsymbol{l}_{1}}(\boldsymbol r ) | \sum_l \phi_{l}^{d}(\boldsymbol r) | \psi_{\boldsymbol{l}_{2}}(\boldsymbol r) \rangle$ \cite{ermin_prl,gunnar_carbon}. 
The coupling elements are calculated by exploiting the nearest-neighbor tight binding approach \cite{ermin_cm,Kochbuch,Kira2006}. \\
The molecules are attached non-covalently to the TMD surface within the distance $R_z$ (Fig. \ref{schema}(a)), leaving the electronic wave function of the TMD unchanged \cite{Hirsch2005}. We can treat the molecule-TMD interaction as a static field induced by the molecular dipole $\bf d$:
\begin{equation}\label{dipolfield} 
 \phi_{l}^{d}(\boldsymbol r) = \frac{ e_{0}}{4\pi\epsilon_{0}} \frac{\boldsymbol {d \cdot (r -R_{l})}}{|\boldsymbol{r-R_{l}}|^{3}} 
\end{equation}
 where $\boldsymbol{R}_{l}$ denotes the position of molecule $l$ with respect to the substrate \cite{ermin_prl}. The dipole field is characterized by a dipole vector 
\begin{equation}
\boldsymbol{d} = d
(\cos(\phi_d)\cos(\alpha_d),
\sin(\phi_d)\cos(\alpha_d),
 \sin(\alpha_d) 
)
\end{equation}
with orientation $\alpha_d$ with respect to xy plane of the TMD, orientation $\phi_d$ in xy direction and dipole moment $d$ (Fig. \ref{schema} (a)). 
Here, we exemplary study the case of a self assembled monolayer of molecules which is characterized by a periodic distribution with lattice constants $\Delta R_{x(y)}$, cf. Fig. \ref{schema}(b).  Furthermore, we take only spin-conserving processes with $\sigma_1=\sigma_2$ into account. With this we find the molecule-TMD coupling (MTC) element as
\begin{eqnarray}\label{MTCFormel}
g_{\boldsymbol{l_{1}l_{2}}}^{\lambda_1 \lambda_2} = \frac{ie_{0}}{2\pi\epsilon_{0}} \sum_{nm} n_{x}n_{y}\delta_{| l_{1x}-l_{2x}|, \frac{2\pi n}{\Delta R_{x}}}
 \delta_{| l_{1y}-l_{2y}|, \frac{2\pi m}{\Delta R_{y}}} \times  \notag \\
 \sum_{j}C_{j}^{\lambda_1 *} (\boldsymbol l_{1}) C_{j}^{\lambda_2} (\boldsymbol l_{2}) \delta_{\boldsymbol{l_{1}-l_{2}},\boldsymbol{q}}\int d\boldsymbol{q}\frac{\boldsymbol d \cdot \boldsymbol{q}}{|\boldsymbol q|^{2}} e^{-R_{z}q_z} 
\end{eqnarray}
with the electric field constant $\epsilon_0$, the molecular coverage $n_{x(y)}=\frac{1}{\Delta R_{x}(y)}$ and tight-binding coefficients $C_j$. 
The periodic molecular lattice allows for well defined momentum transfers determined by the molecular lattice constant, see the appearing Kronecker deltas in Eq. \eqref{MTCFormel}. Hence the MTC element is discrete for periodic distributions, cf. Fig. \ref{kopplungselemente}(a), and decreases with the momentum $\bf q$. 
Note that, for randomly distributed molecules MTC allows continuous momentum transfer which however does not qualitatively change the results presented in this study.

To get access to excitonic properties in TMDs which have been shown to dominate the optical spectra \cite{gunnar_prb}, we transform our system into center of mass $\bf{Q}$ and relative $\bf{q}$ coordinates, i.e. $\bf{Q=k_{2}-k_{1}} $ and ${\bf q}=\alpha{\bf k_{1}}+\beta{\bf k_{2}}$ with $\alpha=\frac{m_{h}}{m_{h}+m_{e}}$ and $\beta=\frac{m_{e}}{m_{h}+m_{e}}$ and project the microscopic polarization into the excitonic basis
\begin{equation}\label{separation}
 p_{\bf {k_{1}k_{2}}}^{\lambda_1\sigma_1 \lambda_2\sigma_2} \rightarrow p_{\bf{qQ}}^{\lambda_1\sigma_1 \lambda_2\sigma_2}= \sum_{\mu} \varphi_{\boldsymbol q}^{\mu} p_{\boldsymbol Q}^{\mu}
\end{equation}
with excitonic state $\mu$, excitonic wave function $ \varphi_{\boldsymbol q}^{\mu}$ and excitonic polarization $p_{\boldsymbol Q}^{\mu}$.
Note that from now on we only consider the energetically lowest $A_{\rm{1s}}$ transitions \cite{THeinz}. Hence it is convenient to denote the excitonic state by the valley, i.e.  $\mu=(K,K')$. The separation ansatz from Eq. \eqref{separation} enables us to decouple the relative from center of mass motion. For the relative part we solve the eigenvalue problem
\begin{equation}
 \frac{\hbar^2 q^{2}}{2m_{red}^{\mu}}  \varphi_{\bf q}^{\mu} 
 - \sum_{\bf k} V_{\text{exc}}(\bf k)  \varphi_{\bf {q-k} }^{\mu}=\varepsilon_{\mu}\varphi_{\bf {q}}^{\mu}
\end{equation}
which is known as the Wannier equation \cite{Kochbuch,Kira2006,kuhn2004}
with excitonic eigenfunctions $\varphi_{\bf q}^{\mu}$ and eigenenergies  $\varepsilon_{\mu}$ in the excitonic state $\mu$ and the reduced mass $m_{red}^{\mu}=\frac{m_h^{\mu}+m_e^{\mu}}{m_h^{\mu}  \cdot m_e^{\mu}}$ . The electron-hole contribution of the Coulomb interaction includes the Fourier transformed Keldysh potential \cite{Keldysh1978}

\begin{equation}
V({\bf k})=\frac{e_0^2}{\epsilon_0 (\epsilon_1 + \epsilon_2) L^2} \frac{1}{|{\bf k}|(1+r_0|{\bf k}|)}, 
\end{equation}
with the sample size $L^2$, dielectric screening constants of the surrounded media $\epsilon_{1,2}$ (in this work free-standing TMD with $\epsilon_{1}=\epsilon_{2}=1$ ) and screening length $r_0=\frac{d\epsilon_{\perp}}{\epsilon_1 + \epsilon_2}$. The thickness $d$ of the material is assumed to be the distance between two sulfur atoms in z direction ($d\approx 0.318$ nm \cite{berkelbach}) and for the dielectric tensor of the TMD layer we assume only the in-plane component $\epsilon_{\perp}$ of the corresponding bulk material \cite{berkelbach}. \\
The excitonic basis approach enables us to write down the corresponding TMD Bloch equation for excitonic microscopic polarization $p_{\bf Q}^{\mu}$
\begin{eqnarray}\label{BWG}
i\hslash \frac{d}{dt} 
p_{\bf{Q}}^{\mu}(t)
&=\left[
\varepsilon_{\mu} - 
\frac{\hbar^2 Q^{2}}{2M} -i\gamma
\right]
p_{\bf{Q}}^{\mu}(t)
+
\Omega(t) \delta_{\bf{Q,0}} \notag \\
& + \sum_{\bf{k}}
G_{\bf{Qk}}^{\mu} p_{\bf{Q-k}}^{\mu} (t)
+
X_{\bf{Q}}^{\mu\nu} p_{\bf{Q}}^{\nu} (t) .
\end{eqnarray}
Here, the Rabi frequency $
\Omega(t)=\frac{ie_{0}\hslash}{m}  
\sum_{\bf{q}} \varphi_{\bf q}^{\mu *} 
{\bf M}_{\bf {q}}^{c\uparrow v\uparrow}\cdot \bf A(t)
$
 originates from the external electromagnetic field $\bf A(t)$ and drives polarization $p_{\bf 0}^{\mu}$ which correspond to excitons with no center of mass momentum. Moreover we introduced a dephasing constant $\gamma$, including dephasing due to  higher order correlations.\\

 The molecule-TMD coupling in excitonic basis reads
\begin{eqnarray}\label{MTCExcitonic}
G_{\bf{Qk}}^{\mu}&=&   \sum_{\bf{q}} \left[
\varphi_{\bf {q}}^{\mu*} g^{cc}_{{\bf q}-\alpha{\bf Q}, {\bf q}-\alpha{\bf Q} + {\bf k}} \varphi_{{\bf q}+\beta {\bf k}}^{\mu} - \right. \notag\\
&&\left. \quad \quad
\varphi_{\bf {q}}^{\mu*} g^{vv}_{{\bf q}+\beta{\bf Q} - {\bf k}, {\bf q}+ \beta {\bf Q}} \varphi_{{\bf q} -\alpha {\bf k}}^{\mu} \right]
\end{eqnarray} 
and drives the indirect polarization $p_{\bf Q}^{\mu}$ and hereby enables a momentum transfer between the molecules and the TMD monolayer. Interestingly, this momentum gives rise to a coherent oscillation transfer  to the other valley $\nu$ via the intervalley coupling (IVC) element 
\begin{eqnarray}\label{IVCFormel}
X_{\bf{Q}}^{\mu\nu}&=&\sum_{\bf qk}
V(Q)
 \frac{\hslash^{2}}{V_{c}^{2} m^{2} \prod\limits_{\sigma}(\omega^{v\sigma}-\omega^{c\sigma})} \times \notag \\ 
 &&(\boldsymbol M_{{\bf q}+\beta {\bf Q}}^{v\uparrow c\uparrow}\cdot {\bf Q})({\bf M}_{{\bf q}-\alpha \bf Q-k}^{c\downarrow v\downarrow}\cdot {\bf Q})
 \varphi_{\bf {q}}^{ \mu*} \cdot
  \varphi_{\bf {q-k}}^{\nu}
\end{eqnarray}  
which is linear in $\mathbf Q$ and vanishes with the center of mass momentum, e.g. $X_{{\bf Q}\rightarrow0}^{\mu\nu}\rightarrow 0$.\\
Figure \ref{kopplungselemente} illustrates the momentum dependent coupling elements both for the molecule-TMD coupling from Eq. \eqref{MTCExcitonic} and intervalley coupling from Eq. \eqref{IVCFormel}. 
In contrast to MTC element which decreases with large center of mass momenta $\mathbf Q$, 
the IVC element is zero for Q=0 and increases linearly with Q (Fig. \ref{kopplungselemente}(b)). Therefore only excitonic coherence exhibiting a center of mass momentum enable intervalley coupling processes, as schematically illustrated in Figure \ref{schema}(c) for optical excitation of the K valley, i.e. $\mu=K$ and $\nu=K'$. The external light excites direct transitions $p_{\bf Q}^K $ in the K valley. 
The optically induced coherence $p_{ \bf 0}^{K}$ now couples to the dipoles of the adsorbed molecules and gains non-zero center of mass momentum resulting in  $p_{\bf Q \neq 0}^K \neq 0$. The gained center of mass momentum opens a channel for coherent intervalley oscillation transfer via $X$ (cf. Eq. \eqref{IVCFormel}) which directly couples  $p^{K}_{\bf Q {\neq 0}}$ and $p^{K'}_{\bf Q {\neq 0}}$.

\begin{figure}[t]
 \begin{center}
\includegraphics[width=\linewidth]{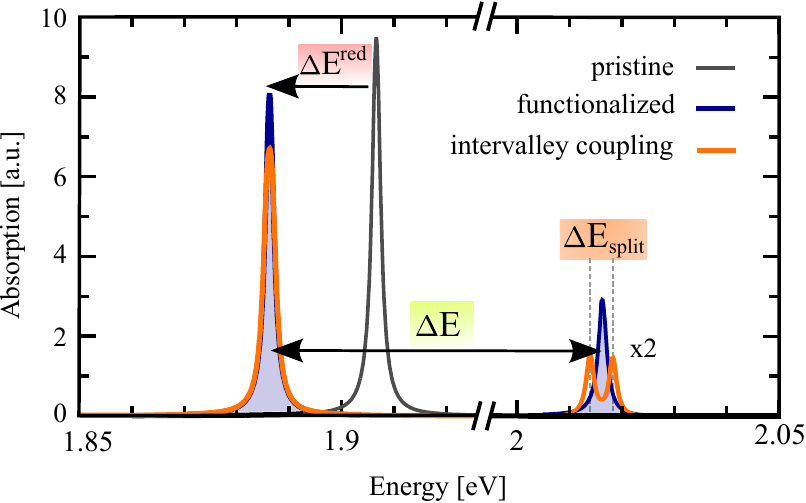}
 \end{center}
 \caption{Excitonic absorption spectrum of pristine and merocyanine-functionalized MoS$_2$. 
For the pristine spectrum (dark gray line) we observe a resonance at 1.9 eV in good agreement with previous studies \cite{gunnar_prb}. We find in case of functionalized MoS$_{2}$ (blue line) that the main peak is redshifted by 20 meV. We also observe an upcoming side peak in approx. 130 meV distance to the main resonance. Taking intervalley coupling into account, cf. orange line,  we observe a further splitting of the side peak in the range of 5 meV. }
 \label{absorption}
\end{figure}

\section{Excitonic absorption spectra}
\begin{figure}[t]
  \begin{center}
   \includegraphics[width=\linewidth]{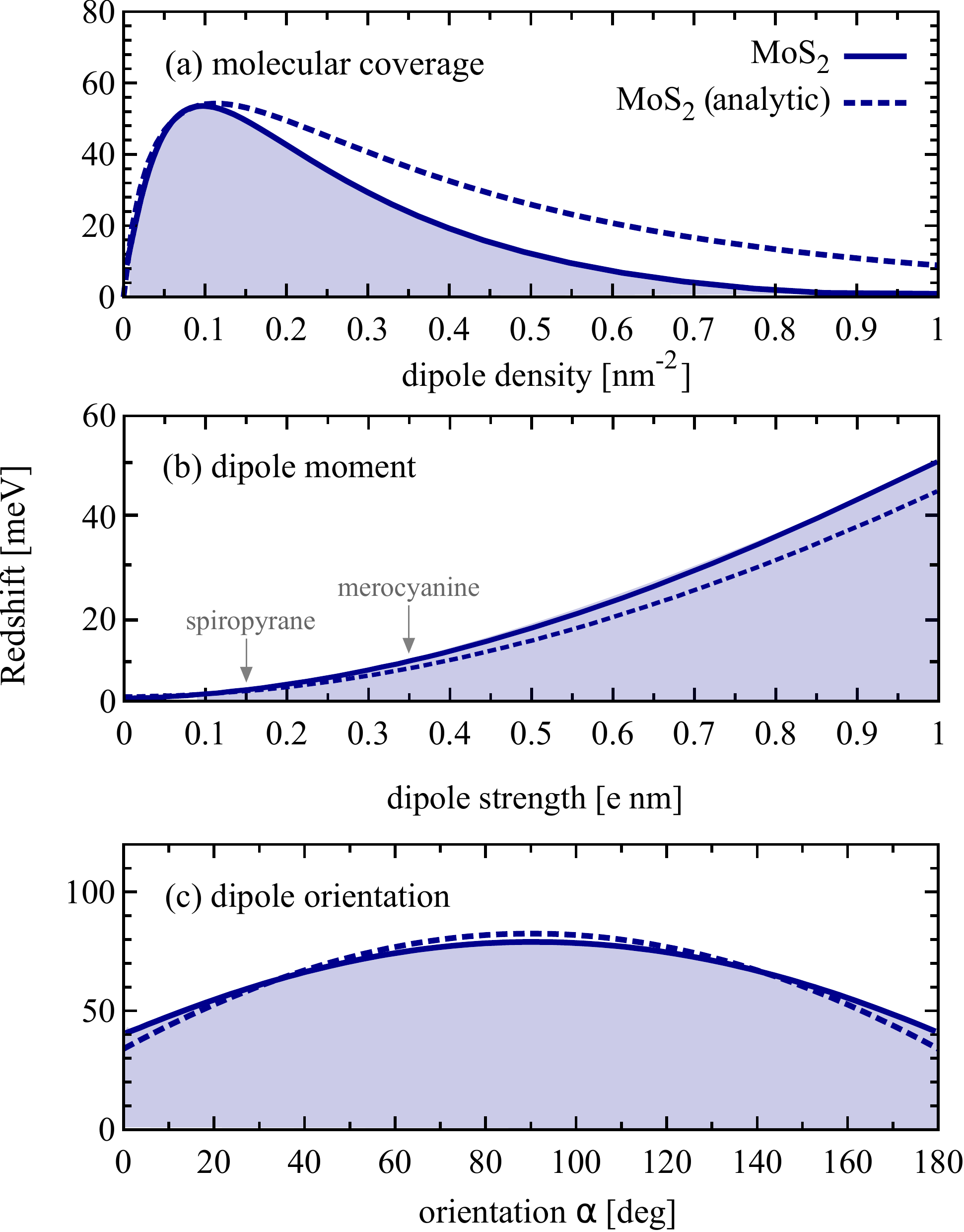}
  \end{center}
  \caption{Redshift of the excitonic absorption spectrum depending on the functionalization parameters (a) molecular coverage, (b) dipole moment, and (c) dipole orientation. The comparison between analytic (numeric) solutions for MoS$_2$, indicated by the blue (dashed blue) lines, shows that the analytic formula reproduces the trends very well.}
 \label{parameter}
\end{figure} 
Now we have all ingredients to calculate the optical response by inserting the solutions of the TMD Bloch equation from Eq. \eqref{BWG} in the Elliott formula Eq. \eqref{Elliott}. The calculated excitonic absorption spectrum, both for pristine (dark gray line) and functionalized (blue line) MoS$_2$ is shown in Fig. \ref{absorption}. We see that the attached merocyanine molecules induce major changes to the spectrum, in particular a redshift $ \Delta \text{E}^{\text{red}} $ of the main peak and an additional side peak. 
In the following we investigate the observed effects in detail, in particular studying the dependence on molecular characteristics. 
As the redshift is the most pronounced feature in the optical absorption and experimentally easy accessible, we focus our analysis on that. Note, however that our calculation revealed similar trends 
for the peak splitting and the intensity ratio between main and side peak. \\
We will first neglect the influence of intervalley coupling and thereby find an analytic solution for the excitonic polarization. \\
Under the assumption that indirect transitions $p_{\bf{Q}-\bf{k}}^{\mu}$ are mainly driven by center of mass momentum $\bf{Q}=|\bf{k}|$, one can find a solution for the direct transitions  
\begin{equation}\label{Pana}
p_{\bf{0}}^{K}(\omega) = \dfrac{\Omega(\omega)}{\hbar\omega-\varepsilon_K - \sum_{{\bf  k}} \dfrac{G_{{\bf 0k}}^K G_{\bf kk}^K}{ \hbar\omega - \varepsilon_K -\frac{\hbar^2 k^2}{2M} }}
\end{equation} 
and we expect resonances in the absorption spectrum to appear at
 \begin{equation}\label{anaFormel}
\varepsilon_{1,2}  =  \varepsilon_\mu + \frac{\hbar^2 Q_{xy}^2}{4M} \mp
\sqrt{
\left(\frac{\hbar^2 Q_{xy}^2}{4M} \right)^2 + \tilde{G}({\bf Q}_{xy})}
\end{equation}
where we evaluated the sum over $\bf k$ in Eq. \eqref{Pana} by focusing on the dominant terms, i.e. $k_x=\frac{2\pi}{\Delta R_x}$ and  $k_y=0$ resulting in $ {Q}_{x}=\frac{2\pi}{\Delta R_x} $ and $Q_{y}=0$. Here, we have introduced the abbreviations 
$\tilde{G}({\bf Q}_{xy})=G_{0, \frac{2\pi}{\Delta R_x} } G_{\frac{2\pi}{\Delta R_x}, \frac{2\pi}{\Delta R_x} } $. \\
For vanishing molecule-TMD coupling, i.e. $\tilde{G}=0$ this reproduces the pristine peak at $\varepsilon_{1}=\varepsilon_{2}=\varepsilon_\mu$. The solutions $\varepsilon_{1,2}$ correspond to the position of the main peak ($\varepsilon_{1}$) and the appearing side peak ($\varepsilon_{2}$) in the functionalized absorption spectra. Figure  \ref{parameter} compares the analytic approach (dashed blue line) from Eq. \eqref{anaFormel} with the exact numerical solution (solid line) with respect to the observed redshift of the main excitonic resonance demonstrating that the analytic solution reproduces well the qualitative trends.
Moreover, Eq.\eqref{anaFormel} reveals that the position of the resonances crucially depends on both the molecular characteristics, such as molecular coverage $n$, dipole moment $d$ and dipole orientation $\alpha$, and on the electronic properties of the TMD entering through the total mass $M$. 

Figure \ref{parameter} shows the dependence of the redshift on the molecular coverage, dipole moment and dipole orientation. 
First, the redshift (Fig. \ref{parameter}(a)) exhibits a maximum for the molecular coverage $n\approx$ 0.1 nm$^{-2}$ which can be explained by the structure of the MTC element, cf. Eq. \ref{MTCFormel}. Projecting the MTC element into the excitonic basis and using the relation $n \propto \frac{Q^2}{2\pi}$ we find $G_{\boldsymbol Q} \propto n^2  e^{-2\sqrt{n}} $ which shows a local maximum as there are two counteracting processes: increasing the number of molecules naturally (i) increases the strength of the coupling ($\propto n^{2}$) as the molecules come closer together, but on the other hand (ii) the impact of the exponential decrease stemming from $q$-dependence in Eq. \eqref{MTCFormel} suppresses the coupling for growing densities resulting in a maximal redshift for $n\approx$ 0.1 nm$^{-2}$.\\
For the dipole moment (Fig. \ref{parameter}(b)) we see a clear quadratic behavior  which is easily explained by  $G_{\boldsymbol Qk}^{\mu} \propto d^2$ stemming from Eq. \eqref{MTCFormel}. The stronger the dipole moment, the stronger the induced dipole field and hence the molecule-TMD coupling. Moreover, we observe that the redshift for the spiropyrane molecule configuration is significantly smaller than for  the merocyanine form due to the much smaller dipole moment. 
Finally, we find for the molecules orientation, $G_{\boldsymbol Qk}^{\mu} \propto \sin^2\alpha - \cos^2\alpha$, which shows a maximal impact for perpendicular orientation.  The dipole axis faces the TMD surface, and similar to classical dipole fields, the induced electric field is maximal in the direction of the dipole axis. 
Moreover, as the induced field in the plane perpendicular to the dipole axis is isotropic to a large extent, we set $\phi_d = 0$ as it does not change the molecule-TMD interaction significantly. \\
Finally, Eq. \eqref{anaFormel} shows that the electronic band structure of the TMDs is a crucial parameter for the position of the resonances. The curvature of the material enters here via the total mass $M$  and we find that the redshift is more pronounced the smaller the total mass is which is in agreement with previous studies on the impact of electronic structure on molecule-substrate coupling \cite{gunnar_carbon}.

\section{Intervalley coupling}
\begin{figure*}[t]
  \begin{center}
    \includegraphics[width=\linewidth]{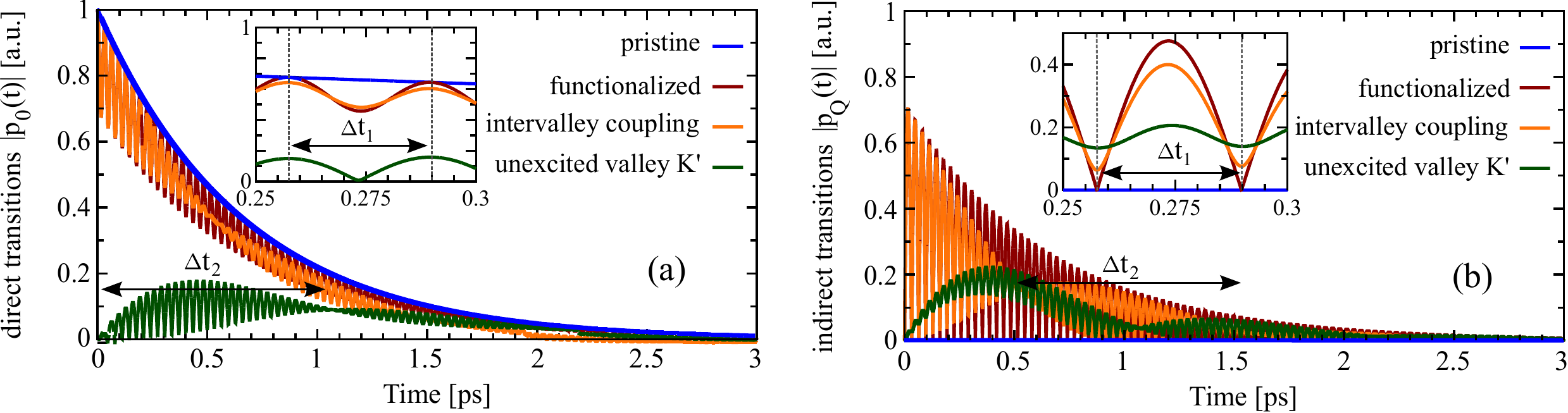}
  \end{center}
  \caption{Excitonic coherence reflecting (a) direct ($|p_{0}^{\mu}|$) and (b) indirect ($|p_{Q}^{\mu}|$) electronic transitions for pristine (blue) and functionalized MoS$_2$, with (orange) and without (red) intervalley coupling.
(a)The polarization $p_0^{K}$ is driven by an external electromagnetic field at t=0. 
In the presence of the molecule we observe a higher and lower frequency oscillation, with oscillation time $\Delta t_1$ and $\Delta t_2$, respectively. The faster oscillation corresponds to the intravalley molecule-induced peak splitting $\Delta \text{E}$ observed in the absorption spectrum in Fig. \ref{absorption}, the slower to the intervalley coupling induced side peak splitting $\Delta \text{E}_{\text{split}}$. 
}
 \label{polarization}
\end{figure*}
 Now we show that the molecule-induced center of mass momentum enables F{\"o}rster like intervalley coupling that is often referred to in literature as Coulomb exchange coupling \cite{gunnar_nano, XiaoPRL2012, Zhu2014,Hongyi_2014NCommValley2014, Wang2015,Hualing2012, Yu2014, Korn2016ncomm, Dery2016}.
Beside the clearly visible molecule induced redshift and peak splitting of the excitonic resonance we observe a further peak splitting $\Delta \text{E}_{\text{split}}$ in the absorption spectrum if we consider intervalley coupling, cf. the orange line in Fig. \ref{absorption}. 
Interestingly, only the side peak shows a splitting whereas the main peak is only decreasing in intensity. This is explained by direct and indirect coupling processes.
As the intervalley coupling element X vanishes for Q=0 (cf. Eq. \eqref{IVCFormel}), the Coulomb exchange interaction can not directly couple the polarization $p_{0}^{K}$ in the K valley to polarization $p_{0}^{K'}$, cf. Eq.\eqref{BWG}.  However, this F\"orster-like term directly couples
 $p_{Q\neq0}^{K}$ to $p_{Q\neq0}^{K'}$ inducing a splitting of their resonances once the molecule ``activates'' those transitions, cf. Fig. \ref{schema}(c). The intervalley coupling is a secondary process which requires the molecule-TMD interaction to work. Hence, controlling the dipole moment of the adsorbed molecules also means to control the X-term induced coherent intervalley coupling mechanism.
  \\
Figure \ref{polarization} shows the directly and indirectly induced polarization in the optically excited and unexcited K' valley.
In Fig. \ref{polarization}(a) we see the squared direct transitions $|p_0|$ in the K valley. Adding the merocyanine molecules to the system (dark red line) the polarization starts to oscillate with the oscillation time $\Delta t_1 = 32$ fs (cf. the inset). This corresponds to $\Delta \rm{E}=130$ meV and represents the difference between the main and the side peak due to the molecule. Furthermore, taking into account intervalley coupling (orange line), one observes an additional oscillation with  $\Delta t_2 = 1.1$ ps. This represents the intervalley peak splitting in the range of $\
\Delta \rm{E_{split}}=4$ meV.
 Finally, the induced coherent intervalley transfer leads to a optically bright polarization $p_0^{K'}$ in the K' valley  (green line) with same oscillation time $\Delta t_{1,2}$. In case of spiropyrane molecule the induced polarization $p_0^{K'}$ is neglicable small. \\ 
Fig. \ref{polarization}(b) shows the directly coupled coherence $|p_Q^{\mu}|$ with center of mass momentum $Q\neq 0$ in the K and K' valley. Naturally, $|p_Q^{\mu}|$ is zero for the pristine system (blue line) since only transitions $Q=0$ are optically driven due to the vanishing momentum transfer of the driving electromagnetic field. Adding molecules (dark red line) enables indirect transitions with $Q\neq 0$ and therefore polarization $|p_Q|$ is induced. Again, we observe an oscillation with oscillation time $\Delta t_1$ due to MTC and an exponential decrease due to dephasing. Also for the indirect transitions the intervalley coupling (orange line) leads to a further oscillation $\Delta t_2$ and one observes a pronounced polarization in the K' valley (green line). Since $p_Q^{K'}$ couples directly to $p_Q^{K}$ the oscillation transfer between these resonant quantities is extremely efficient and hence both oscillate with the same amplitude. This explains the side peak splitting in the absorption spectrum that can be interpreted as the response to induced indirect transitions $p_Q^{K}$. Even though the induced intervalley coupling can be only indirectly measured in linear optics, the possibility to control intervalley coupling shows exciting prospects in the non-linear regime. \\

\section{Conclusion}
Using a microscopic and semi-analytic approach we have studied the optical properties of TMD-hybrid materials in the linear regime. We have shown that non-covalently attached molecules exhibiting a dipole moment significantly influence the optical response of the energetically lowest excitonic transitions. In particular, our calculations predict a redshift of tens of meV and a peak splitting in the range of 100 meV.  
The optical properties of the functionalized material strongly depend on the molecules characteristics, such as molecular coverage, dipole orientation and dipole moment. 
We further have shown that the Coulomb-induced intervalley coupling can be turned on in the presence of molecules. 
This allows to control the coherent intervalley coupling process and may open new possibilities in tailoring hybrid materials for spin- and valleytronics application. 

\section{Acknowledgment}
This project has received funding from the European Union's Horizon 2020 research and innovation programme under grant agreement No 696656 within the Graphene Flagship (EM,GB) and the Swedish Research Council (EM,GB). Furthermore, we acknowledge support by  the Chalmers Area of Advance in Nanoscience and Nanotechnology (MF,EM) and by the Deutsche Forschungsgemeinschaft through SFB 658 (EM) and 951 (AK, MF, GB).  A.K.  acknowledges the Research and Innovation Staff Exchange program SONAR (Marie Curie Actions) of the European Union’s Horizon 2020 [Grant number 734690].\\


\begin{thebibliography}{33}%
\makeatletter
\providecommand \@ifxundefined [1]{%
 \@ifx{#1\undefined}
}%
\providecommand \@ifnum [1]{%
 \ifnum #1\expandafter \@firstoftwo
 \else \expandafter \@secondoftwo
 \fi
}%
\providecommand \@ifx [1]{%
 \ifx #1\expandafter \@firstoftwo
 \else \expandafter \@secondoftwo
 \fi
}%
\providecommand \natexlab [1]{#1}%
\providecommand \enquote  [1]{``#1''}%
\providecommand \bibnamefont  [1]{#1}%
\providecommand \bibfnamefont [1]{#1}%
\providecommand \citenamefont [1]{#1}%
\providecommand \href@noop [0]{\@secondoftwo}%
\providecommand \href [0]{\begingroup \@sanitize@url \@href}%
\providecommand \@href[1]{\@@startlink{#1}\@@href}%
\providecommand \@@href[1]{\endgroup#1\@@endlink}%
\providecommand \@sanitize@url [0]{\catcode `\\12\catcode `\$12\catcode
  `\&12\catcode `\#12\catcode `\^12\catcode `\_12\catcode `\%12\relax}%
\providecommand \@@startlink[1]{}%
\providecommand \@@endlink[0]{}%
\providecommand \url  [0]{\begingroup\@sanitize@url \@url }%
\providecommand \@url [1]{\endgroup\@href {#1}{\urlprefix }}%
\providecommand \urlprefix  [0]{URL }%
\providecommand \Eprint [0]{\href }%
\providecommand \doibase [0]{http://dx.doi.org/}%
\providecommand \selectlanguage [0]{\@gobble}%
\providecommand \bibinfo  [0]{\@secondoftwo}%
\providecommand \bibfield  [0]{\@secondoftwo}%
\providecommand \translation [1]{[#1]}%
\providecommand \BibitemOpen [0]{}%
\providecommand \bibitemStop [0]{}%
\providecommand \bibitemNoStop [0]{.\EOS\space}%
\providecommand \EOS [0]{\spacefactor3000\relax}%
\providecommand \BibitemShut  [1]{\csname bibitem#1\endcsname}%
\let\auto@bib@innerbib\@empty
\bibitem [{\citenamefont {Bergh\"auser}\ and\ \citenamefont
  {Malic}(2013)}]{pss_gunnar}%
  \BibitemOpen
  \bibfield  {author} {\bibinfo {author} {\bibfnamefont {G.}~\bibnamefont
  {Bergh\"auser}}\ and\ \bibinfo {author} {\bibfnamefont {E.}~\bibnamefont
  {Malic}},\ }\href {\doibase 10.1002/pssb.201300181} {\bibfield  {journal}
  {\bibinfo  {journal} {physica status solidi (b)}\ }\textbf {\bibinfo {volume}
  {250}},\ \bibinfo {pages} {2678} (\bibinfo {year} {2013})}\BibitemShut
  {NoStop}%
\bibitem [{\citenamefont {Bergh\"auser}\ and\ \citenamefont
  {Malic}(2014{\natexlab{a}})}]{gunnar_carbon}%
  \BibitemOpen
  \bibfield  {author} {\bibinfo {author} {\bibfnamefont {G.}~\bibnamefont
  {Bergh\"auser}}\ and\ \bibinfo {author} {\bibfnamefont {E.}~\bibnamefont
  {Malic}},\ }\href {\doibase http://dx.doi.org/10.1016/j.carbon.2013.12.063}
  {\bibfield  {journal} {\bibinfo  {journal} {Carbon}\ }\textbf {\bibinfo
  {volume} {69}},\ \bibinfo {pages} {536 } (\bibinfo {year}
  {2014}{\natexlab{a}})}\BibitemShut {NoStop}%
\bibitem [{\citenamefont {Malic}\ \emph {et~al.}(2012)\citenamefont {Malic},
  \citenamefont {Setaro}, \citenamefont {Bluemmel}, \citenamefont
  {Sanz-Navarro}, \citenamefont {Ordejón}, \citenamefont {Reich},\ and\
  \citenamefont {Knorr}}]{ermin_cm}%
  \BibitemOpen
  \bibfield  {author} {\bibinfo {author} {\bibfnamefont {E.}~\bibnamefont
  {Malic}}, \bibinfo {author} {\bibfnamefont {A.}~\bibnamefont {Setaro}},
  \bibinfo {author} {\bibfnamefont {P.}~\bibnamefont {Bluemmel}}, \bibinfo
  {author} {\bibfnamefont {C.~F.}\ \bibnamefont {Sanz-Navarro}}, \bibinfo
  {author} {\bibfnamefont {P.}~\bibnamefont {Ordejón}}, \bibinfo {author}
  {\bibfnamefont {S.}~\bibnamefont {Reich}}, \ and\ \bibinfo {author}
  {\bibfnamefont {A.}~\bibnamefont {Knorr}},\ }\href
  {http://stacks.iop.org/0953-8984/24/i=39/a=394006} {\bibfield  {journal}
  {\bibinfo  {journal} {Journal of Physics: Condensed Matter}\ }\textbf
  {\bibinfo {volume} {24}},\ \bibinfo {pages} {394006} (\bibinfo {year}
  {2012})}\BibitemShut {NoStop}%
\bibitem [{\citenamefont {Malic}\ \emph {et~al.}(2011)\citenamefont {Malic},
  \citenamefont {Weber}, \citenamefont {Richter}, \citenamefont {Atalla},
  \citenamefont {Klamroth}, \citenamefont {Saalfrank}, \citenamefont {Reich},\
  and\ \citenamefont {Knorr}}]{ermin_prl}%
  \BibitemOpen
  \bibfield  {author} {\bibinfo {author} {\bibfnamefont {E.}~\bibnamefont
  {Malic}}, \bibinfo {author} {\bibfnamefont {C.}~\bibnamefont {Weber}},
  \bibinfo {author} {\bibfnamefont {M.}~\bibnamefont {Richter}}, \bibinfo
  {author} {\bibfnamefont {V.}~\bibnamefont {Atalla}}, \bibinfo {author}
  {\bibfnamefont {T.}~\bibnamefont {Klamroth}}, \bibinfo {author}
  {\bibfnamefont {P.}~\bibnamefont {Saalfrank}}, \bibinfo {author}
  {\bibfnamefont {S.}~\bibnamefont {Reich}}, \ and\ \bibinfo {author}
  {\bibfnamefont {A.}~\bibnamefont {Knorr}},\ }\href {\doibase
  10.1103/PhysRevLett.106.097401} {\bibfield  {journal} {\bibinfo  {journal}
  {Phys. Rev. Lett.}\ }\textbf {\bibinfo {volume} {106}},\ \bibinfo {pages}
  {097401} (\bibinfo {year} {2011})}\BibitemShut {NoStop}%
\bibitem [{\citenamefont {Malic}\ \emph {et~al.}(2014)\citenamefont {Malic},
  \citenamefont {Appel}, \citenamefont {Hofmann},\ and\ \citenamefont
  {Rubio}}]{malic2014forster}%
  \BibitemOpen
  \bibfield  {author} {\bibinfo {author} {\bibfnamefont {E.}~\bibnamefont
  {Malic}}, \bibinfo {author} {\bibfnamefont {H.}~\bibnamefont {Appel}},
  \bibinfo {author} {\bibfnamefont {O.~T.}\ \bibnamefont {Hofmann}}, \ and\
  \bibinfo {author} {\bibfnamefont {A.}~\bibnamefont {Rubio}},\ }\href@noop {}
  {\bibfield  {journal} {\bibinfo  {journal} {The journal of physical
  chemistry. C, Nanomaterials and interfaces}\ }\textbf {\bibinfo {volume}
  {118}},\ \bibinfo {pages} {9283} (\bibinfo {year} {2014})}\BibitemShut
  {NoStop}%
\bibitem [{\citenamefont {Butler}\ \emph {et~al.}(2013)\citenamefont {Butler},
  \citenamefont {Hollen}, \citenamefont {Cao}, \citenamefont {Cui},
  \citenamefont {Gupta}, \citenamefont {Gutiérrez}, \citenamefont {Heinz},
  \citenamefont {Hong}, \citenamefont {Huang}, \citenamefont {Ismach},
  \citenamefont {Johnston-Halperin}, \citenamefont {Kuno}, \citenamefont
  {Plashnitsa}, \citenamefont {Robinson}, \citenamefont {Ruoff}, \citenamefont
  {Salahuddin}, \citenamefont {Shan}, \citenamefont {Shi}, \citenamefont
  {Spencer}, \citenamefont {Terrones}, \citenamefont {Windl},\ and\
  \citenamefont {Goldberger}}]{Butler2013}%
  \BibitemOpen
  \bibfield  {author} {\bibinfo {author} {\bibfnamefont {S.~Z.}\ \bibnamefont
  {Butler}}, \bibinfo {author} {\bibfnamefont {S.~M.}\ \bibnamefont {Hollen}},
  \bibinfo {author} {\bibfnamefont {L.}~\bibnamefont {Cao}}, \bibinfo {author}
  {\bibfnamefont {Y.}~\bibnamefont {Cui}}, \bibinfo {author} {\bibfnamefont
  {J.~A.}\ \bibnamefont {Gupta}}, \bibinfo {author} {\bibfnamefont {H.~R.}\
  \bibnamefont {Gutiérrez}}, \bibinfo {author} {\bibfnamefont {T.~F.}\
  \bibnamefont {Heinz}}, \bibinfo {author} {\bibfnamefont {S.~S.}\ \bibnamefont
  {Hong}}, \bibinfo {author} {\bibfnamefont {J.}~\bibnamefont {Huang}},
  \bibinfo {author} {\bibfnamefont {A.~F.}\ \bibnamefont {Ismach}}, \bibinfo
  {author} {\bibfnamefont {E.}~\bibnamefont {Johnston-Halperin}}, \bibinfo
  {author} {\bibfnamefont {M.}~\bibnamefont {Kuno}}, \bibinfo {author}
  {\bibfnamefont {V.~V.}\ \bibnamefont {Plashnitsa}}, \bibinfo {author}
  {\bibfnamefont {R.~D.}\ \bibnamefont {Robinson}}, \bibinfo {author}
  {\bibfnamefont {R.~S.}\ \bibnamefont {Ruoff}}, \bibinfo {author}
  {\bibfnamefont {S.}~\bibnamefont {Salahuddin}}, \bibinfo {author}
  {\bibfnamefont {J.}~\bibnamefont {Shan}}, \bibinfo {author} {\bibfnamefont
  {L.}~\bibnamefont {Shi}}, \bibinfo {author} {\bibfnamefont {M.~G.}\
  \bibnamefont {Spencer}}, \bibinfo {author} {\bibfnamefont {M.}~\bibnamefont
  {Terrones}}, \bibinfo {author} {\bibfnamefont {W.}~\bibnamefont {Windl}}, \
  and\ \bibinfo {author} {\bibfnamefont {J.~E.}\ \bibnamefont {Goldberger}},\
  }\href {\doibase 10.1021/nn400280c} {\bibfield  {journal} {\bibinfo
  {journal} {ACS Nano}\ }\textbf {\bibinfo {volume} {7}},\ \bibinfo {pages}
  {2898} (\bibinfo {year} {2013})},\ \bibinfo {note} {pMID: 23464873},\ \Eprint
  {http://arxiv.org/abs/http://dx.doi.org/10.1021/nn400280c}
  {http://dx.doi.org/10.1021/nn400280c} \BibitemShut {NoStop}%
\bibitem [{\citenamefont {Haug}\ and\ \citenamefont {Koch}()}]{Kochbuch}%
  \BibitemOpen
  \bibfield  {author} {\bibinfo {author} {\bibfnamefont {H.}~\bibnamefont
  {Haug}}\ and\ \bibinfo {author} {\bibfnamefont {S.~W.}\ \bibnamefont
  {Koch}},\ }\href@noop {} {\emph {\bibinfo {title} {Quantum Theory of the
  Optical and Electronic Properties of Semiconductors}}}\ (\bibinfo
  {publisher} {5th ed. (World Scientific Publishing Co. Pre. Ltd., Singapore,
  2004).})\BibitemShut {NoStop}%
\bibitem [{\citenamefont {Kormányos}\ \emph {et~al.}(2015)\citenamefont
  {Kormányos}, \citenamefont {Burkard}, \citenamefont {Gmitra}, \citenamefont
  {Fabian}, \citenamefont {Zólyomi}, \citenamefont {Drummond},\ and\
  \citenamefont {Fal’ko}}]{andor}%
  \BibitemOpen
  \bibfield  {author} {\bibinfo {author} {\bibfnamefont {A.}~\bibnamefont
  {Kormányos}}, \bibinfo {author} {\bibfnamefont {G.}~\bibnamefont {Burkard}},
  \bibinfo {author} {\bibfnamefont {M.}~\bibnamefont {Gmitra}}, \bibinfo
  {author} {\bibfnamefont {J.}~\bibnamefont {Fabian}}, \bibinfo {author}
  {\bibfnamefont {V.}~\bibnamefont {Zólyomi}}, \bibinfo {author}
  {\bibfnamefont {N.~D.}\ \bibnamefont {Drummond}}, \ and\ \bibinfo {author}
  {\bibfnamefont {V.}~\bibnamefont {Fal’ko}},\ }\href
  {http://stacks.iop.org/2053-1583/2/i=2/a=022001} {\bibfield  {journal}
  {\bibinfo  {journal} {2D Materials}\ }\textbf {\bibinfo {volume} {2}},\
  \bibinfo {pages} {022001} (\bibinfo {year} {2015})}\BibitemShut {NoStop}%
\bibitem [{\citenamefont {Mak}\ \emph {et~al.}(2010)\citenamefont {Mak},
  \citenamefont {Lee}, \citenamefont {Hone}, \citenamefont {Shan},\ and\
  \citenamefont {Heinz}}]{THeinz}%
  \BibitemOpen
  \bibfield  {author} {\bibinfo {author} {\bibfnamefont {K.~F.}\ \bibnamefont
  {Mak}}, \bibinfo {author} {\bibfnamefont {C.}~\bibnamefont {Lee}}, \bibinfo
  {author} {\bibfnamefont {J.}~\bibnamefont {Hone}}, \bibinfo {author}
  {\bibfnamefont {J.}~\bibnamefont {Shan}}, \ and\ \bibinfo {author}
  {\bibfnamefont {T.~F.}\ \bibnamefont {Heinz}},\ }\href {\doibase
  10.1103/PhysRevLett.105.136805} {\bibfield  {journal} {\bibinfo  {journal}
  {Phys. Rev. Lett.}\ }\textbf {\bibinfo {volume} {105}},\ \bibinfo {pages}
  {136805} (\bibinfo {year} {2010})}\BibitemShut {NoStop}%
\bibitem [{\citenamefont {{Cao Ting}}\ \emph {et~al.}(2012)\citenamefont {{Cao
  Ting}}, \citenamefont {{Wang Gang}}, \citenamefont {{Han Wenpeng}},
  \citenamefont {{Ye Huiqi}}, \citenamefont {{Zhu Chuanrui}}, \citenamefont
  {{Shi Junren}}, \citenamefont {{Niu Qian}}, \citenamefont {{Tan Pingheng}},
  \citenamefont {{Wang Enge}}, \citenamefont {{Liu Baoli}},\ and\ \citenamefont
  {{Feng Ji}}}]{Cao2012}%
  \BibitemOpen
  \bibfield  {author} {\bibinfo {author} {\bibnamefont {{Cao Ting}}}, \bibinfo
  {author} {\bibnamefont {{Wang Gang}}}, \bibinfo {author} {\bibnamefont {{Han
  Wenpeng}}}, \bibinfo {author} {\bibnamefont {{Ye Huiqi}}}, \bibinfo {author}
  {\bibnamefont {{Zhu Chuanrui}}}, \bibinfo {author} {\bibnamefont {{Shi
  Junren}}}, \bibinfo {author} {\bibnamefont {{Niu Qian}}}, \bibinfo {author}
  {\bibnamefont {{Tan Pingheng}}}, \bibinfo {author} {\bibnamefont {{Wang
  Enge}}}, \bibinfo {author} {\bibnamefont {{Liu Baoli}}}, \ and\ \bibinfo
  {author} {\bibnamefont {{Feng Ji}}},\ }\href {\doibase
  http://dx.doi.org/10.1038/ncomms1882} {\bibfield  {journal} {\bibinfo
  {journal} {Nat Commun}\ }\textbf {\bibinfo {volume} {3}},\ \bibinfo {pages}
  {887} (\bibinfo {year} {2012})},\ \bibinfo {note}
  {10.1038/ncomms1882}\BibitemShut {NoStop}%
\bibitem [{\citenamefont {Yao}\ \emph {et~al.}(2008)\citenamefont {Yao},
  \citenamefont {Xiao},\ and\ \citenamefont {Niu}}]{valley_pola}%
  \BibitemOpen
  \bibfield  {author} {\bibinfo {author} {\bibfnamefont {W.}~\bibnamefont
  {Yao}}, \bibinfo {author} {\bibfnamefont {D.}~\bibnamefont {Xiao}}, \ and\
  \bibinfo {author} {\bibfnamefont {Q.}~\bibnamefont {Niu}},\ }\href {\doibase
  10.1103/PhysRevB.77.235406} {\bibfield  {journal} {\bibinfo  {journal} {Phys.
  Rev. B}\ }\textbf {\bibinfo {volume} {77}},\ \bibinfo {pages} {235406}
  (\bibinfo {year} {2008})}\BibitemShut {NoStop}%
\bibitem [{\citenamefont {Lu}\ \emph {et~al.}(2013)\citenamefont {Lu},
  \citenamefont {Yao}, \citenamefont {Xiao},\ and\ \citenamefont
  {Shen}}]{valley_pola2}%
  \BibitemOpen
  \bibfield  {author} {\bibinfo {author} {\bibfnamefont {H.-Z.}\ \bibnamefont
  {Lu}}, \bibinfo {author} {\bibfnamefont {W.}~\bibnamefont {Yao}}, \bibinfo
  {author} {\bibfnamefont {D.}~\bibnamefont {Xiao}}, \ and\ \bibinfo {author}
  {\bibfnamefont {S.-Q.}\ \bibnamefont {Shen}},\ }\href {\doibase
  10.1103/PhysRevLett.110.016806} {\bibfield  {journal} {\bibinfo  {journal}
  {Phys. Rev. Lett.}\ }\textbf {\bibinfo {volume} {110}},\ \bibinfo {pages}
  {016806} (\bibinfo {year} {2013})}\BibitemShut {NoStop}%
\bibitem [{\citenamefont {Fai}\ \emph {et~al.}(2013)\citenamefont {Fai},
  \citenamefont {Keliang}, \citenamefont {Changgu}, \citenamefont {Hyoung},
  \citenamefont {James}, \citenamefont {F.},\ and\ \citenamefont
  {Jie}}]{RIS_0}%
  \BibitemOpen
  \bibfield  {author} {\bibinfo {author} {\bibfnamefont {M.~K.}\ \bibnamefont
  {Fai}}, \bibinfo {author} {\bibfnamefont {H.}~\bibnamefont {Keliang}},
  \bibinfo {author} {\bibfnamefont {L.}~\bibnamefont {Changgu}}, \bibinfo
  {author} {\bibfnamefont {L.~G.}\ \bibnamefont {Hyoung}}, \bibinfo {author}
  {\bibfnamefont {H.}~\bibnamefont {James}}, \bibinfo {author} {\bibfnamefont
  {H.~T.}\ \bibnamefont {F.}}, \ and\ \bibinfo {author} {\bibfnamefont
  {S.}~\bibnamefont {Jie}},\ }\href {\doibase
  http://dx.doi.org/10.1038/nmat3505} {\bibfield  {journal} {\bibinfo
  {journal} {Nat Mater}\ }\textbf {\bibinfo {volume} {12}},\ \bibinfo {pages}
  {207–211} (\bibinfo {year} {2013})}\BibitemShut {NoStop}%
\bibitem [{\citenamefont {Bergh\"auser}\ and\ \citenamefont
  {Malic}(2014{\natexlab{b}})}]{gunnar_prb}%
  \BibitemOpen
  \bibfield  {author} {\bibinfo {author} {\bibfnamefont {G.}~\bibnamefont
  {Bergh\"auser}}\ and\ \bibinfo {author} {\bibfnamefont {E.}~\bibnamefont
  {Malic}},\ }\href {\doibase 10.1103/PhysRevB.89.125309} {\bibfield  {journal}
  {\bibinfo  {journal} {Phys. Rev. B}\ }\textbf {\bibinfo {volume} {89}},\
  \bibinfo {pages} {125309} (\bibinfo {year} {2014}{\natexlab{b}})}\BibitemShut
  {NoStop}%
\bibitem [{\citenamefont {Baker}\ \emph {et~al.}(1952)\citenamefont {Baker},
  \citenamefont {Tompkins}, \citenamefont {Fahim}, \citenamefont {Fleifel},
  \citenamefont {Bergmann}, \citenamefont {Kalmus}, \citenamefont {Fischer},
  \citenamefont {Hirshberg}, \citenamefont {Arnstein}, \citenamefont {Ward},
  \citenamefont {Day}, \citenamefont {Bradley}, \citenamefont {Tadros},
  \citenamefont {Kamel}, \citenamefont {Bailey}, \citenamefont {Bates},
  \citenamefont {Ing}, \citenamefont {Warne}, \citenamefont {Neale},
  \citenamefont {Williams}, \citenamefont {Henbest}, \citenamefont {Sharpe},
  \citenamefont {Lamberton}, \citenamefont {Hart}, \citenamefont {Bunton},
  \citenamefont {Halevi}, \citenamefont {Thurston}, \citenamefont {Walker},
  \citenamefont {Robinson}, \citenamefont {Mann}, \citenamefont {Smith},
  \citenamefont {Hammick}, \citenamefont {Roe}, \citenamefont {Peat},
  \citenamefont {Whelan},\ and\ \citenamefont {Thomas}}]{FisherSpiro}%
  \BibitemOpen
  \bibfield  {author} {\bibinfo {author} {\bibfnamefont {E.~H.}\ \bibnamefont
  {Baker}}, \bibinfo {author} {\bibfnamefont {F.~C.}\ \bibnamefont {Tompkins}},
  \bibinfo {author} {\bibfnamefont {H.~A.}\ \bibnamefont {Fahim}}, \bibinfo
  {author} {\bibfnamefont {A.~M.}\ \bibnamefont {Fleifel}}, \bibinfo {author}
  {\bibfnamefont {F.}~\bibnamefont {Bergmann}}, \bibinfo {author}
  {\bibfnamefont {A.}~\bibnamefont {Kalmus}}, \bibinfo {author} {\bibfnamefont
  {E.}~\bibnamefont {Fischer}}, \bibinfo {author} {\bibfnamefont
  {Y.}~\bibnamefont {Hirshberg}}, \bibinfo {author} {\bibfnamefont {H.~R.~V.}\
  \bibnamefont {Arnstein}}, \bibinfo {author} {\bibfnamefont {E.~R.}\
  \bibnamefont {Ward}}, \bibinfo {author} {\bibfnamefont {L.~A.}\ \bibnamefont
  {Day}}, \bibinfo {author} {\bibfnamefont {R.~S.}\ \bibnamefont {Bradley}},
  \bibinfo {author} {\bibfnamefont {W.}~\bibnamefont {Tadros}}, \bibinfo
  {author} {\bibfnamefont {M.}~\bibnamefont {Kamel}}, \bibinfo {author}
  {\bibfnamefont {A.~S.}\ \bibnamefont {Bailey}}, \bibinfo {author}
  {\bibfnamefont {D.~H.}\ \bibnamefont {Bates}}, \bibinfo {author}
  {\bibfnamefont {H.~R.}\ \bibnamefont {Ing}}, \bibinfo {author} {\bibfnamefont
  {M.~A.}\ \bibnamefont {Warne}}, \bibinfo {author} {\bibfnamefont
  {E.}~\bibnamefont {Neale}}, \bibinfo {author} {\bibfnamefont {L.~T.~D.}\
  \bibnamefont {Williams}}, \bibinfo {author} {\bibfnamefont {H.~B.}\
  \bibnamefont {Henbest}}, \bibinfo {author} {\bibfnamefont {A.~G.}\
  \bibnamefont {Sharpe}}, \bibinfo {author} {\bibfnamefont {A.~H.}\
  \bibnamefont {Lamberton}}, \bibinfo {author} {\bibfnamefont {E.~P.}\
  \bibnamefont {Hart}}, \bibinfo {author} {\bibfnamefont {C.~A.}\ \bibnamefont
  {Bunton}}, \bibinfo {author} {\bibfnamefont {E.~A.}\ \bibnamefont {Halevi}},
  \bibinfo {author} {\bibfnamefont {J.~P.}\ \bibnamefont {Thurston}}, \bibinfo
  {author} {\bibfnamefont {J.}~\bibnamefont {Walker}}, \bibinfo {author}
  {\bibfnamefont {R.~A.}\ \bibnamefont {Robinson}}, \bibinfo {author}
  {\bibfnamefont {F.~G.}\ \bibnamefont {Mann}}, \bibinfo {author}
  {\bibfnamefont {B.~B.}\ \bibnamefont {Smith}}, \bibinfo {author}
  {\bibfnamefont {D.~L.}\ \bibnamefont {Hammick}}, \bibinfo {author}
  {\bibfnamefont {A.~M.}\ \bibnamefont {Roe}}, \bibinfo {author} {\bibfnamefont
  {S.}~\bibnamefont {Peat}}, \bibinfo {author} {\bibfnamefont {W.~J.}\
  \bibnamefont {Whelan}}, \ and\ \bibinfo {author} {\bibfnamefont {G.~J.}\
  \bibnamefont {Thomas}},\ }\href {\doibase 10.1039/JR9520004518} {\bibfield
  {journal} {\bibinfo  {journal} {J. Chem. Soc.}\ ,\ \bibinfo {pages} {4518}}
  (\bibinfo {year} {1952})}\BibitemShut {NoStop}%
\bibitem [{\citenamefont {Feierabend}\ \emph {et~al.}(2017)\citenamefont
  {Feierabend}, \citenamefont {Bergh{\"a}user}, \citenamefont {Knorr},\ and\
  \citenamefont {Malic}}]{maja_sensor}%
  \BibitemOpen
  \bibfield  {author} {\bibinfo {author} {\bibfnamefont {M.}~\bibnamefont
  {Feierabend}}, \bibinfo {author} {\bibfnamefont {G.}~\bibnamefont
  {Bergh{\"a}user}}, \bibinfo {author} {\bibfnamefont {A.}~\bibnamefont
  {Knorr}}, \ and\ \bibinfo {author} {\bibfnamefont {E.}~\bibnamefont
  {Malic}},\ }\href@noop {} {\bibfield  {journal} {\bibinfo  {journal} {Nature
  Communications}\ }\textbf {\bibinfo {volume} {8}},\ \bibinfo {pages} {14776}
  (\bibinfo {year} {2017})}\BibitemShut {NoStop}%
\bibitem [{\citenamefont {Malic}\ and\ \citenamefont
  {Knorr}(2013)}]{carbonbuch}%
  \BibitemOpen
  \bibfield  {author} {\bibinfo {author} {\bibfnamefont {E.}~\bibnamefont
  {Malic}}\ and\ \bibinfo {author} {\bibfnamefont {A.}~\bibnamefont {Knorr}},\
  }\href@noop {} {\emph {\bibinfo {title} {Graphene and Carbon Nanotubes:
  Ultrafast Optics and Relaxation Dynamics}}}\ (\bibinfo  {publisher} {John
  Wiley \& Sons},\ \bibinfo {year} {2013})\BibitemShut {NoStop}%
\bibitem [{\citenamefont {Xiao}\ \emph
  {et~al.}(2012{\natexlab{a}})\citenamefont {Xiao}, \citenamefont {Liu},
  \citenamefont {Feng}, \citenamefont {Xu},\ and\ \citenamefont {Yao}}]{Xiao}%
  \BibitemOpen
  \bibfield  {author} {\bibinfo {author} {\bibfnamefont {D.}~\bibnamefont
  {Xiao}}, \bibinfo {author} {\bibfnamefont {G.-B.}\ \bibnamefont {Liu}},
  \bibinfo {author} {\bibfnamefont {W.}~\bibnamefont {Feng}}, \bibinfo {author}
  {\bibfnamefont {X.}~\bibnamefont {Xu}}, \ and\ \bibinfo {author}
  {\bibfnamefont {W.}~\bibnamefont {Yao}},\ }\href {\doibase
  10.1103/PhysRevLett.108.196802} {\bibfield  {journal} {\bibinfo  {journal}
  {Phys. Rev. Lett.}\ }\textbf {\bibinfo {volume} {108}},\ \bibinfo {pages}
  {196802} (\bibinfo {year} {2012}{\natexlab{a}})}\BibitemShut {NoStop}%
\bibitem [{\citenamefont {Ochoa}\ and\ \citenamefont {Rold\'an}(2013)}]{ochoa}%
  \BibitemOpen
  \bibfield  {author} {\bibinfo {author} {\bibfnamefont {H.}~\bibnamefont
  {Ochoa}}\ and\ \bibinfo {author} {\bibfnamefont {R.}~\bibnamefont
  {Rold\'an}},\ }\href {\doibase 10.1103/PhysRevB.87.245421} {\bibfield
  {journal} {\bibinfo  {journal} {Phys. Rev. B}\ }\textbf {\bibinfo {volume}
  {87}},\ \bibinfo {pages} {245421} (\bibinfo {year} {2013})}\BibitemShut
  {NoStop}%
\bibitem [{\citenamefont {Kira}\ and\ \citenamefont {Koch}(2006)}]{Kira2006}%
  \BibitemOpen
  \bibfield  {author} {\bibinfo {author} {\bibfnamefont {M.}~\bibnamefont
  {Kira}}\ and\ \bibinfo {author} {\bibfnamefont {S.}~\bibnamefont {Koch}},\
  }\href {\doibase http://dx.doi.org/10.1016/j.pquantelec.2006.12.002}
  {\bibfield  {journal} {\bibinfo  {journal} {Progress in Quantum Electronics}\
  }\textbf {\bibinfo {volume} {30}},\ \bibinfo {pages} {155 } (\bibinfo {year}
  {2006})}\BibitemShut {NoStop}%
\bibitem [{\citenamefont {Hirsch}\ and\ \citenamefont
  {Vostrowsky}(2005)}]{Hirsch2005}%
  \BibitemOpen
  \bibfield  {author} {\bibinfo {author} {\bibfnamefont {A.}~\bibnamefont
  {Hirsch}}\ and\ \bibinfo {author} {\bibfnamefont {O.}~\bibnamefont
  {Vostrowsky}},\ }\href {\doibase 10.1007/b98169} {\emph {\bibinfo {title}
  {Functional Molecular Nanostructures}}}\ (\bibinfo  {publisher} {Springer
  Berlin Heidelberg},\ \bibinfo {address} {Berlin, Heidelberg},\ \bibinfo
  {year} {2005})\BibitemShut {NoStop}%
\bibitem [{\citenamefont {Axt}\ and\ \citenamefont {Kuhn}(2004)}]{kuhn2004}%
  \BibitemOpen
  \bibfield  {author} {\bibinfo {author} {\bibfnamefont {V.}~\bibnamefont
  {Axt}}\ and\ \bibinfo {author} {\bibfnamefont {T.}~\bibnamefont {Kuhn}},\
  }\href@noop {} {\bibfield  {journal} {\bibinfo  {journal} {Reports on
  Progress in Physics}\ }\textbf {\bibinfo {volume} {67}},\ \bibinfo {pages}
  {433} (\bibinfo {year} {2004})}\BibitemShut {NoStop}%
\bibitem [{\citenamefont {Keldysh}(1978)}]{Keldysh1978}%
  \BibitemOpen
  \bibfield  {author} {\bibinfo {author} {\bibnamefont {Keldysh}},\ }\href@noop
  {} {\bibfield  {journal} {\bibinfo  {journal} {JETP Lett.}\ }\textbf
  {\bibinfo {volume} {29}},\ \bibinfo {pages} {658} (\bibinfo {year}
  {1978})}\BibitemShut {NoStop}%
\bibitem [{\citenamefont {Berkelbach}\ \emph {et~al.}(2013)\citenamefont
  {Berkelbach}, \citenamefont {Hybertsen},\ and\ \citenamefont
  {Reichman}}]{berkelbach}%
  \BibitemOpen
  \bibfield  {author} {\bibinfo {author} {\bibfnamefont {T.~C.}\ \bibnamefont
  {Berkelbach}}, \bibinfo {author} {\bibfnamefont {M.~S.}\ \bibnamefont
  {Hybertsen}}, \ and\ \bibinfo {author} {\bibfnamefont {D.~R.}\ \bibnamefont
  {Reichman}},\ }\href {\doibase 10.1103/PhysRevB.88.045318} {\bibfield
  {journal} {\bibinfo  {journal} {Phys. Rev. B}\ }\textbf {\bibinfo {volume}
  {88}},\ \bibinfo {pages} {045318} (\bibinfo {year} {2013})}\BibitemShut
  {NoStop}%
\bibitem [{\citenamefont {Schmidt}\ \emph {et~al.}(2016)\citenamefont
  {Schmidt}, \citenamefont {Bergh�user}, \citenamefont {Schneider},
  \citenamefont {Selig}, \citenamefont {Tonndorf}, \citenamefont {Malic},
  \citenamefont {Knorr}, \citenamefont {Michaelis~de Vasconcellos},\ and\
  \citenamefont {Bratschitsch}}]{gunnar_nano}%
  \BibitemOpen
  \bibfield  {author} {\bibinfo {author} {\bibfnamefont {R.}~\bibnamefont
  {Schmidt}}, \bibinfo {author} {\bibfnamefont {G.}~\bibnamefont
  {Bergh�user}}, \bibinfo {author} {\bibfnamefont {R.}~\bibnamefont
  {Schneider}}, \bibinfo {author} {\bibfnamefont {M.}~\bibnamefont {Selig}},
  \bibinfo {author} {\bibfnamefont {P.}~\bibnamefont {Tonndorf}}, \bibinfo
  {author} {\bibfnamefont {E.}~\bibnamefont {Malic}}, \bibinfo {author}
  {\bibfnamefont {A.}~\bibnamefont {Knorr}}, \bibinfo {author} {\bibfnamefont
  {S.}~\bibnamefont {Michaelis~de Vasconcellos}}, \ and\ \bibinfo {author}
  {\bibfnamefont {R.}~\bibnamefont {Bratschitsch}},\ }\href@noop {} {\bibfield
  {journal} {\bibinfo  {journal} {Nano letters}\ }\textbf {\bibinfo {volume}
  {16}},\ \bibinfo {pages} {2945} (\bibinfo {year} {2016})}\BibitemShut
  {NoStop}%
\bibitem [{\citenamefont {Xiao}\ \emph
  {et~al.}(2012{\natexlab{b}})\citenamefont {Xiao}, \citenamefont {Liu},
  \citenamefont {Feng}, \citenamefont {Xu},\ and\ \citenamefont
  {Yao}}]{XiaoPRL2012}%
  \BibitemOpen
  \bibfield  {author} {\bibinfo {author} {\bibfnamefont {D.}~\bibnamefont
  {Xiao}}, \bibinfo {author} {\bibfnamefont {G.-B.}\ \bibnamefont {Liu}},
  \bibinfo {author} {\bibfnamefont {W.}~\bibnamefont {Feng}}, \bibinfo {author}
  {\bibfnamefont {X.}~\bibnamefont {Xu}}, \ and\ \bibinfo {author}
  {\bibfnamefont {W.}~\bibnamefont {Yao}},\ }\href {\doibase
  10.1103/PhysRevLett.108.196802} {\bibfield  {journal} {\bibinfo  {journal}
  {Phys. Rev. Lett.}\ }\textbf {\bibinfo {volume} {108}},\ \bibinfo {pages}
  {196802} (\bibinfo {year} {2012}{\natexlab{b}})}\BibitemShut {NoStop}%
\bibitem [{\citenamefont {Zhu}\ \emph {et~al.}(2014)\citenamefont {Zhu},
  \citenamefont {Zhang}, \citenamefont {Glazov}, \citenamefont {Urbaszek},
  \citenamefont {Amand}, \citenamefont {Ji}, \citenamefont {Liu},\ and\
  \citenamefont {Marie}}]{Zhu2014}%
  \BibitemOpen
  \bibfield  {author} {\bibinfo {author} {\bibfnamefont {C.~R.}\ \bibnamefont
  {Zhu}}, \bibinfo {author} {\bibfnamefont {K.}~\bibnamefont {Zhang}}, \bibinfo
  {author} {\bibfnamefont {M.}~\bibnamefont {Glazov}}, \bibinfo {author}
  {\bibfnamefont {B.}~\bibnamefont {Urbaszek}}, \bibinfo {author}
  {\bibfnamefont {T.}~\bibnamefont {Amand}}, \bibinfo {author} {\bibfnamefont
  {Z.~W.}\ \bibnamefont {Ji}}, \bibinfo {author} {\bibfnamefont {B.~L.}\
  \bibnamefont {Liu}}, \ and\ \bibinfo {author} {\bibfnamefont
  {X.}~\bibnamefont {Marie}},\ }\href {\doibase 10.1103/PhysRevB.90.161302}
  {\bibfield  {journal} {\bibinfo  {journal} {Phys. Rev. B}\ }\textbf {\bibinfo
  {volume} {90}},\ \bibinfo {pages} {161302} (\bibinfo {year}
  {2014})}\BibitemShut {NoStop}%
\bibitem [{\citenamefont {Yu}\ \emph {et~al.}(2014)\citenamefont {Yu},
  \citenamefont {Liu}, \citenamefont {Gong}, \citenamefont {Xu},\ and\
  \citenamefont {Yao}}]{Hongyi_2014NCommValley2014}%
  \BibitemOpen
  \bibfield  {author} {\bibinfo {author} {\bibfnamefont {H.}~\bibnamefont
  {Yu}}, \bibinfo {author} {\bibfnamefont {G.-B.}\ \bibnamefont {Liu}},
  \bibinfo {author} {\bibfnamefont {P.}~\bibnamefont {Gong}}, \bibinfo {author}
  {\bibfnamefont {X.}~\bibnamefont {Xu}}, \ and\ \bibinfo {author}
  {\bibfnamefont {W.}~\bibnamefont {Yao}},\ }\href
  {http://dx.doi.org/10.1038/ncomms4876} {\bibfield  {journal} {\bibinfo
  {journal} {Nature Communications}\ }\textbf {\bibinfo {volume} {5}},\
  \bibinfo {pages} {3876} (\bibinfo {year} {2014})}\BibitemShut {NoStop}%
\bibitem [{\citenamefont {Wang}\ \emph {et~al.}(2015)\citenamefont {Wang},
  \citenamefont {Luo}, \citenamefont {Yabushita}, \citenamefont {Wu},
  \citenamefont {Kobayashi}, \citenamefont {Chen},\ and\ \citenamefont
  {Li}}]{Wang2015}%
  \BibitemOpen
  \bibfield  {author} {\bibinfo {author} {\bibfnamefont {Y.-T.}\ \bibnamefont
  {Wang}}, \bibinfo {author} {\bibfnamefont {C.-W.}\ \bibnamefont {Luo}},
  \bibinfo {author} {\bibfnamefont {A.}~\bibnamefont {Yabushita}}, \bibinfo
  {author} {\bibfnamefont {K.-H.}\ \bibnamefont {Wu}}, \bibinfo {author}
  {\bibfnamefont {T.}~\bibnamefont {Kobayashi}}, \bibinfo {author}
  {\bibfnamefont {C.-H.}\ \bibnamefont {Chen}}, \ and\ \bibinfo {author}
  {\bibfnamefont {L.-J.}\ \bibnamefont {Li}},\ }\href
  {http://dx.doi.org/10.1038/srep08289} {\bibfield  {journal} {\bibinfo
  {journal} {Scientific Reports}\ }\textbf {\bibinfo {volume} {5}},\ \bibinfo
  {pages} {8289} (\bibinfo {year} {2015})}\BibitemShut {NoStop}%
\bibitem [{\citenamefont {Zeng}\ \emph {et~al.}(2012)\citenamefont {Zeng},
  \citenamefont {Dai}, \citenamefont {Yao}, \citenamefont {Xiao},\ and\
  \citenamefont {Cui}}]{Hualing2012}%
  \BibitemOpen
  \bibfield  {author} {\bibinfo {author} {\bibfnamefont {H.}~\bibnamefont
  {Zeng}}, \bibinfo {author} {\bibfnamefont {J.}~\bibnamefont {Dai}}, \bibinfo
  {author} {\bibfnamefont {W.}~\bibnamefont {Yao}}, \bibinfo {author}
  {\bibfnamefont {D.}~\bibnamefont {Xiao}}, \ and\ \bibinfo {author}
  {\bibfnamefont {X.}~\bibnamefont {Cui}},\ }\href {\doibase
  10.1038/nnano.2012.95} {\bibfield  {journal} {\bibinfo  {journal} {Nat Nano}\
  }\textbf {\bibinfo {volume} {7}},\ \bibinfo {pages} {490} (\bibinfo {year}
  {2012})}\BibitemShut {NoStop}%
\bibitem [{\citenamefont {Yu}\ and\ \citenamefont {Wu}(2014)}]{Yu2014}%
  \BibitemOpen
  \bibfield  {author} {\bibinfo {author} {\bibfnamefont {T.}~\bibnamefont
  {Yu}}\ and\ \bibinfo {author} {\bibfnamefont {M.~W.}\ \bibnamefont {Wu}},\
  }\href {\doibase 10.1103/PhysRevB.89.205303} {\bibfield  {journal} {\bibinfo
  {journal} {Phys. Rev. B}\ }\textbf {\bibinfo {volume} {89}},\ \bibinfo
  {pages} {205303} (\bibinfo {year} {2014})}\BibitemShut {NoStop}%
\bibitem [{\citenamefont {Plechinger}\ \emph {et~al.}(2016)\citenamefont
  {Plechinger}, \citenamefont {Nagler}, \citenamefont {Arora}, \citenamefont
  {Schmidt}, \citenamefont {Chernikov}, \citenamefont {del {\'A}guila},
  \citenamefont {Christianen}, \citenamefont {Bratschitsch}, \citenamefont
  {Sch{\"u}ller},\ and\ \citenamefont {Korn}}]{Korn2016ncomm}%
  \BibitemOpen
  \bibfield  {author} {\bibinfo {author} {\bibfnamefont {G.}~\bibnamefont
  {Plechinger}}, \bibinfo {author} {\bibfnamefont {P.}~\bibnamefont {Nagler}},
  \bibinfo {author} {\bibfnamefont {A.}~\bibnamefont {Arora}}, \bibinfo
  {author} {\bibfnamefont {R.}~\bibnamefont {Schmidt}}, \bibinfo {author}
  {\bibfnamefont {A.}~\bibnamefont {Chernikov}}, \bibinfo {author}
  {\bibfnamefont {A.~G.}\ \bibnamefont {del {\'A}guila}}, \bibinfo {author}
  {\bibfnamefont {P.~C.~M.}\ \bibnamefont {Christianen}}, \bibinfo {author}
  {\bibfnamefont {R.}~\bibnamefont {Bratschitsch}}, \bibinfo {author}
  {\bibfnamefont {C.}~\bibnamefont {Sch{\"u}ller}}, \ and\ \bibinfo {author}
  {\bibfnamefont {T.}~\bibnamefont {Korn}},\ }\href
  {http://dx.doi.org/10.1038/ncomms12715} {\bibfield  {journal} {\bibinfo
  {journal} {Nature Communications}\ }\textbf {\bibinfo {volume} {7}},\
  \bibinfo {pages} {12715} (\bibinfo {year} {2016})}\BibitemShut {NoStop}%
\bibitem [{\citenamefont {Dery}(2016)}]{Dery2016}%
  \BibitemOpen
  \bibfield  {author} {\bibinfo {author} {\bibfnamefont {H.}~\bibnamefont
  {Dery}},\ }\href {\doibase 10.1103/PhysRevB.94.075421} {\bibfield  {journal}
  {\bibinfo  {journal} {Phys. Rev. B}\ }\textbf {\bibinfo {volume} {94}},\
  \bibinfo {pages} {075421} (\bibinfo {year} {2016})}\BibitemShut {NoStop}%
\end{thebibliography}
\end{document}